\documentclass[a4paper,preprint]{revtex4}
\usepackage{amsmath}
\usepackage{amsfonts}
\usepackage{amssymb}
\usepackage{graphicx}
\usepackage{subscript}
\usepackage{siunitx}
\usepackage{xcolor}
\usepackage{gensymb}

\begin{document}

\title{Generalized inverse patchy colloid model}

\author{Monika Stipsitz}
\affiliation{Institute for Theoretical Physics, Technische Universit\"at 
Wien, Wiedner Hauptstra{\ss}e 8-10, A-1040 Wien, Austria}

\author{Gerhard Kahl}
\email{gerhard.kahl@tuwien.ac.at}
\affiliation{Institute for Theoretical Physics and Center for 
Computational Materials Science (CMS), Technische Universit\"at Wien, 
Wiedner Hauptstra{\ss}e 8-10, A-1040 Wien, Austria}

\author{Emanuela Bianchi}
\email{emanuela.bianchi@tuwien.ac.at}
\affiliation{Institute for Theoretical Physics, Technische Universit\"at 
Wien, Wiedner Hauptstra{\ss}e 8-10, A-1040 Wien, Austria.}

\date{\today}

\begin{abstract} 
We generalize the inverse patchy colloid model that was originally developed for heterogeneously charged particles with two identical polar patches and an oppositely charged equator to a model that can have a considerably richer surface pattern.  Based on a Debye-H\"uckel framework, we propose a coarse-grained description of the effective pair interactions that is applicable to particles with an arbitrary patch decoration. We demonstrate the versatility of this approach by applying it to models with (i) two differently charged and/or sized patches, and (ii) three, possibly different patches.
\end{abstract}

\maketitle

\section{Introduction}

Colloids with patterned surfaces (commonly referred to as patchy particles) are considered as very versatile building entities whose shape and surface decoration can be designed such that they support the self-assembly of target structures with desired properties~\cite{kretzschmar:review,bianchi:review}. The self-organizing behavior of these units is based on their ability to form  directional and highly selective bonds via the specific interaction features of the different surface areas. The versatility of this class of particles can be even enhanced if their surface decoration is characterized by charged regions, since in this case different surface areas can either be mutually attractive or mutually repulsive according to their respective charges. Many naturally occurring units at the nanoscale, such as proteins, virus capsids, and spotted vesicles, are characterized by differently charged moieties~\cite{Daniel_ACS_2010,Zhang_2008,Zhang_2012,Christian_NatMat_2009,Juhl_2006}; in simple model systems an heterogeneously charged surface can be realized in the so-called inverse patchy colloids (IPCs)~\cite{bianchi:ipcfirst}.

IPCs are charged spherical colloids decorated by oppositely charged regions, realized, for instance, by polyelectrolyte stars adsorbed onto the particle surface~\cite{Blaak_2008,Blaak_2012}. Based on the features of the complex units presented in Ref.~\cite{Blaak_2008}, a related model has been developed~\cite{bianchi:ipcfirst}, where two identical, charged patches are located on the poles of an oppositely charged colloid. Closed expressions for the single particle potential and for the pair potential of two interacting IPCs can be written using a standard formalism of electrostatics within the Debye-H{\"u}ckel approach: the resulting analytical description of the pair interactions is based on approximating -- under high screening conditions -- the ratio between spherical Bessel functions of consecutive orders with a Yukawa-like term~\cite{bianchi:ipcfirst}. In order to make IPC systems amenable to theoretical approaches and/or computer simulations a coarse-graining description was also introduced, leading to considerably simplified expressions for the pair interaction between IPCs with two symmetric patches~\cite{bianchi:ipcfirst}.  These interactions, that depend on both the relative distance and the relative orientation of the two IPCs, are able to form bonds in a highly selective manner, leading to complex self-assembly scenarios~\cite{bianchi:2d2013,bianchi:2d2014,noya:planes2014,noya:planes2015,yura:theory,silvano:30n}.

The rapid progress in synthesizing heterogeneously patterned colloids with desired patch decorations~\cite{Kretzschmar_2013,Kretzschmar_2012,Kagome_2011,Kraft_2009,vanOostrum_2015,Ravaine_2014,Ravaine_2014bis} urges to push forward the described IPC model towards a more sophisticated framework where particles can have an arbitrary patch decoration. The present contribution is dedicated to present a general formalism that allows to calculate the effective potential between two arbitrarily decorated IPCs. Similar to the original derivation for two identical patches~\cite{bianchi:ipcfirst}, we replace the differently charged regions of the particle by point charges that are located in their respective centers of charge. For the ensuing charge distribution we solve Maxwell's equations, along with the boundary conditions at the surface, and obtain the potential created by a single IPC: expanding both the charge density and the ansatz for the potential in terms of spherical harmonics we can derive for a particular patch decoration a set of equations for the expansion coefficients. Based on this potential, we can then calculate the effective interaction between two IPCs. The algorithm, although amenable to an arbitrary patch decoration, becomes rapidly involved with an increasing number of patches and/or the complexity of the patch decoration. In this contribution we demonstrate the versatility of our approach by considering in detail the two-patch case with a size and/or charge disparity in the patches; moreover we extend the present approach to IPCs with a symmetric patch decoration of three identical patches.

The manuscript is organized as follows. In section~\ref{sec:dh}, we provide the theoretical description of the effective interactions between two IPCs with polar patches that are characterized by two possibly different sets of parameters. Following the Debye-H{\"uckel} approach, we first -- analytically -- calculate the electrostatic potential around a single two-patch IPC (see subsection~\ref{sec:singlepot}) and then we -- numerically -- determine the effective potential between two IPCs (see subsection~\ref{sec:pairpot}). In section~\ref{sec:coarse-grained}, we introduce the coarse-grained model and we motivate the choice of its parameters with the mapping procedure that links the analytical to the coarse-grained description. In section~\ref{sec:results}, we consider a broad selection of IPC systems and we show results for the Debye-H{\"u}ckel potential (see subsection~\ref{sec:DH}) and for the coarse-grained potential (see subsection~~\ref{sec:CG}). In subsection~\ref{sec:DH}, we distinguish three cases: IPCs with identical (i.e. symmetric) patches as in Refs.~\cite{bianchi:2d2013,bianchi:2d2014}, IPCs with patches carrying different charges, and IPCs with patches having different surface extensions; results for the case of symmetric patches are compared to those obtained with the Yukawa approximation used in Ref.~\cite{bianchi:ipcfirst}. In subsection~\ref{sec:CG}, we consider the case of patches that are asymmetric in charge. Our conclusions are outlined in section~\ref{sec:conclusion}. Finally, in the Appendix we demonstrate how to extend our approach to IPCs with three identical patches. 

\section{The Debye-H\"uckel description}
\label{sec:dh}

We consider an IPC of radius $\sigma$ surrounded by a dielectric solvent.  According to Gauss' law the electrostatic field outside a sphere with a homogeneously distributed surface charge is identical to the field generated by a point charge positioned in the center of the dielectric colloidal particle. When a colloid has a heterogeneously charged surface, the charge of the different areas (either the patches or the bare colloid) can be replaced by a discrete distribution of point charges positioned at the respective centers of charge inside the particle: while the center of charge of the colloid coincides with the center of the colloidal sphere, the centers of charge of the patches are located inside the particle in a well-defined geometry (that exactly reflects the patch decoration), each at a specific distance from the particle center. The inside of the colloid is inaccessible to both the co- and counter-ions of the electrolyte solution, while outside the particle these ions are part of the surrounding medium. For sake of simplicity we assume that the dielectric permittivity, $\varepsilon$, has the same value both inside and outside the colloid.

In Subsection~\ref{sec:singlepot} we derive the screened electrostatic potential generated by a single particle with a discrete distribution of charges: the potential is calculated both inside and outside the colloid by expanding it in terms of spherical harmonics; then, by imposing electrostatic boundary conditions, a set of linear equations for the expansion coefficients is obtained. The system of equations can be numerically solved, leading eventually to the single particle potential.  Based on this information, we numerically calculate in Subsection~\ref{sec:pairpot} the pair potential between two identical particles.

The present approach can be applied to a colloidal particle decorated by an arbitrary number of patches, $n_p$, at well-defined positions. As the algorithm becomes rapidly involved with increasing $n_p$, we focus here on IPCs with two differently charged or sized patches. To demonstrate the versatility of the concept, we briefly outline the case of IPCs with three patches in the Appendix.

\subsection{The electrostatic potential around a single two-patch IPC}\label{sec:singlepot}

\subsubsection{Calculation of the potential inside the colloid}

We describe the heterogeneously charged colloid as a dielectric sphere with a discrete charge distribution $\rho(r,\theta,\varphi)$ located in its interior. For sake of simplicity we set the elementary charge $q_{e}$ to unity. 
The potential generated by the distribution of the charges inside the particle satisfies Poisson's equation (Gaussian units are used in the following)
\begin{equation}\label{eq:poisson}
\Delta\Phi^{(1)}(r,\theta,\varphi) =
\frac{4 \pi}{\varepsilon}\rho(r,\theta,\varphi) ;
\end{equation}
here the charge density, $\rho(r,\theta,\varphi)$, is the result of the central charge of the colloid, $Z_{c}$, and the two out-of-center charges of the patches, $Z_{\rm p_1}$ and $Z_{\rm p_2}$, respectively; the latter ones are
assumed to be positioned at distances $a_{1}$ and $a_{2}$ from the particle center, in directions opposite to each other. In spherical coordinates the charge density is thus
\begin{widetext}
\begin{equation}\label{eq:rho}
\rho (r,\theta,\varphi)= -Z_{c}\delta({\bf r})-
Z_{\rm p_1}\frac{1}{a_1^2}\delta(r-a_1)\delta(\theta-\frac{\pi}{2})\delta(\varphi)-
Z_{\rm p_2}\frac{1}{a_2^2}\delta(r-a_2)\delta(\theta-\frac{\pi}{2})\delta(\varphi-\pi) . 
\end{equation}
\end{widetext}
The general solution of Equation~(\ref{eq:poisson}) is the sum of the general solution of the corresponding homogeneous equation (also known as Laplace's equation) and a particular solution of the inhomogeneous
equation.  The first contribution, $\Phi^{(1)}_{\rm hom}(r,\theta,\varphi)$, is
\begin{equation}\label{eq:phi_in_hom}
\Phi^{(1)}_{\rm hom}(r,\theta,\varphi) =
\sum\limits_{{\ell=0}}^{\infty}\sum\limits_{m=-\ell}^{+\ell}\left[B_{\ell m}r^{\ell}+
C_{\ell m}r^{-\ell-1}\right]Y_{\ell m}(\theta,\varphi) ,
\end{equation}
the $Y_{\ell m}(\theta,\varphi)$ being the spherical harmonics. Since the potential inside the colloidal sphere should be well-behaved at $r=0$, the coefficients $C_{\ell m}$ are set to zero.

A particular solution of Equation~(\ref{eq:poisson}), $\Phi^{(1)}_{\rm part}(r,\theta,\varphi)$, depends on the specific charge distribution $\rho(r,\theta,\varphi)$.  In our case -- Equation~(\ref{eq:rho}) -- we end up with the following expression
\begin{widetext}
\begin{equation}\label{eq:phi_in_part}
\begin{aligned}
\Phi^{(1)}_{\rm part}(r,\theta,\varphi)& =
\frac{4\pi}{\varepsilon}
\sum\limits_{{\ell=0}}^{\infty}\sum\limits_{m=-\ell}^{+\ell}\frac{1}{2\ell+1}\left[\frac{Z_{\rm p_1}~a_{1}^{\ell}}
{r^{\ell+1}}Y_{\ell m}^{*}\left(\frac{\pi}{2},0\right)+
\frac{Z_{\rm p_2}~a_{2}^{\ell}}{r^{\ell+1}}Y_{\ell m}^{*}
\left(\frac{\pi}{2},\pi\right)\right]Y_{\ell m}(\theta,\varphi) \\
&+\frac{4\pi}{\varepsilon}Z_{c}\frac{1}{r}Y_{00}^{*}Y_{00} ,
\end{aligned}
\end{equation}
\end{widetext}
where the star denotes complex conjugation.  We note that for two identical patches, the expansion of the potential in terms of spherical harmonics can be replaced by an expansion in terms of Legendre polynomials~\cite{bianchi:ipcfirst}. Here the coordinate system is fixed such that one patch is located at $\varphi=0$ and $\theta=\frac{\pi}{2}$.  This system is used in all following steps.

The solution inside the colloid, $\Phi^{(1)}(r,\theta,\varphi)$, is finally given by
\begin{equation}
\Phi^{(1)}(r,\theta,\varphi) = \Phi^{(1)}_{\rm hom}(r,\theta,\varphi)
+ \Phi^{(1)}_{\rm part}(r,\theta,\varphi).
\end{equation}

\subsubsection{Calculation of the potential outside the colloid}

Outside the colloid the dielectric solvent has to be taken into account: it consists of a large number of co- and counter-ions which will be treated within a mean field approach such that Boltzmann statistics apply. The characteristic features of the dielectric solvent enter only via the inverse Debye screening length $\kappa$~\cite{dlvo}.  At low ion densities, a linearization of the Poisson-Boltzmann equation leads to the Helmholtz equation for the potential outside the colloid, $\Phi^{(2)}(r,\theta,\varphi)$,
\begin{equation}\label{eq:helmoltz}
\Delta \Phi^{(2)}(r,\theta,\varphi) = \kappa^{2}\Phi^{(2)}(r,\theta,\varphi).
\end{equation}
In its most common form this equation has a different sign and is solved by an expansion in terms of the spherical Bessel functions $j_{\ell}(x)$ and $y_{\ell}(x)$. Using the substitution $\kappa \rightarrow i\kappa$ in the expression for the standard solution, we arrive at the following solution for Equation~(\ref{eq:helmoltz})
\begin{equation}\label{eq:phi_out}
\Phi^{(2)}(r,\theta,\varphi) =
\sum\limits_{{\ell=0}}^{\infty}\sum\limits_{m=-\ell}^{+\ell}\left[A_{\ell m}j_{\ell}(i\kappa r) +
B_{\ell m}y_{\ell}(i\kappa r)\right]Y_{\ell m}(\theta,\varphi),
\end{equation}
where $i$ is the complex unit and
\begin{align}\label{eq:bessel}
&j_{n}(i \kappa r)=(i)^{n}(\kappa r)^{n}\left(\frac{1}{\kappa r}\dfrac{d}{d(\kappa r)}\right)^{n}\frac{\sinh{\kappa r}}{\kappa r} \\
&y_{n}(i \kappa r)=-(i)^{n}(\kappa r)^{n}\left(\frac{1}{\kappa r}\dfrac{d}{d(\kappa r)}\right)^{n}\frac{\cosh{\kappa r}}{\kappa r} , 
\end{align}
with $n = 0, 1, 2, 3, \dots$.  Since $\Phi^{(2)}(r,\theta,\varphi)$ has to be regular for $r\rightarrow \infty$, while $j_{n}(i \kappa r)$ and $y_{n}(i \kappa r)$ both diverge at large distances, we introduce -- following standard procedures -- suitable linear combinations of the two sets of functions, namely
\begin{widetext}
\begin{align}\label{eq:hankel}
&h_{n}^{(1)}(i \kappa r) = j_{n}(i \kappa r) + iy_{n}(i \kappa r) =
-(i)^{n}(\kappa r)^{n}\left(\frac{1}{\kappa r}\dfrac{d}{d(\kappa r)}\right)^{n}\frac{e^{-\kappa r}}{\kappa r} \\
&h_{n}^{(2)}(i \kappa r) = j_{n}(i \kappa r) - iy_{n}(i \kappa r) =
(i)^{n}(\kappa r)^{n}\left(\frac{1}{\kappa r}\dfrac{d}{d(\kappa r)}\right)^{n}\frac{e^{\kappa r}}{\kappa r},
\end{align}
\end{widetext}
which are commonly known as Hankel function of the first and second kind, respectively.  Since the Hankel functions of the second kind do not decay exponentially with the distance, only the Hankel functions of the first kind are retained in the following.

Thus the potential outside the colloid can be written as
\begin{align}
\label{eq:phi_out_bis}
\Phi^{(2)}(r,\theta,\varphi) = 
\sum\limits_{{\ell=0}}^{\infty}\sum\limits_{m=-\ell}^{+\ell} C_{\ell m}h^{(1)}_{\ell}(i\kappa r)Y_{\ell m}(\theta,\varphi).
\end{align}

\subsubsection{Linking the two potentials via boundary conditions}

To solve the system of differential equations, we need to consider the proper boundary conditions. First the potential must vanish at infinity and has to be continuous at $r=\sigma$. Second, in the absence of surface charges, the tangential component of the electrostatic field has to be continuous. Finally, the normal component of the displacement field must be continuous. Since the relative permittivity $\varepsilon$ is taken to be the same inside and outside of the colloid, these boundary conditions can be written as
\begin{align}\label{eq:boundary}
\Phi^{(1)}(r,\theta,\varphi)|_{r=\sigma} &= \Phi^{(2)}(r,\theta,\varphi)|_{r=\sigma} \nonumber \\
\partial_{r}\Phi^{(1)}(r,\theta,\varphi)|_{r=\sigma}  &= \partial_{r}\Phi^{(2)}(r,\theta,\varphi)|_{r=\sigma}~.
\end{align}

We now insert the expansions of $\Phi^{(1)}(r,\theta,\varphi)$ -- see Equations~(\ref{eq:phi_in_part}) and (\ref{eq:phi_in_hom}) -- and of $\Phi^{(2)}(r=\sigma,\theta,\varphi)$ -- see Equation (\ref{eq:phi_out_bis}) -- into these relations. Since the spherical harmonics are linearly independent, we obtain for each index combination $(\ell, m)$ an equation for the corresponding, yet unknown coefficient that has to be solved numerically; the series expansions are truncated at a suitable upper limit, $\ell_{\rm max}$. Once these coefficients are known, the resulting electrostatic potential around a single IPC is thus available. Being interested only in the potential
outside the colloid, we will omit in the following the superscript 2 and refer to the screened electrostatic potential in the region of interest as $\Phi(r,\theta,\varphi)$.

Since in our particular case the patches are arranged on the ($\theta=\frac{\pi}{2}$)-plane, the $\theta$-symmetry is conserved and only spherical harmonics with an even sum ($\ell + m$) contribute to the expansion. If, in addition, the patches are symmetric in size and charge, a further symmetry can be exploited; as a consequence, only functions with $m\bmod n_{p}=0$ need to be considered~\cite{bianchi:ipcfirst}, where $n_{p}$ denotes the number of patches in the system. 

\begin{figure}[!h]
\includegraphics[width=0.5\textwidth]{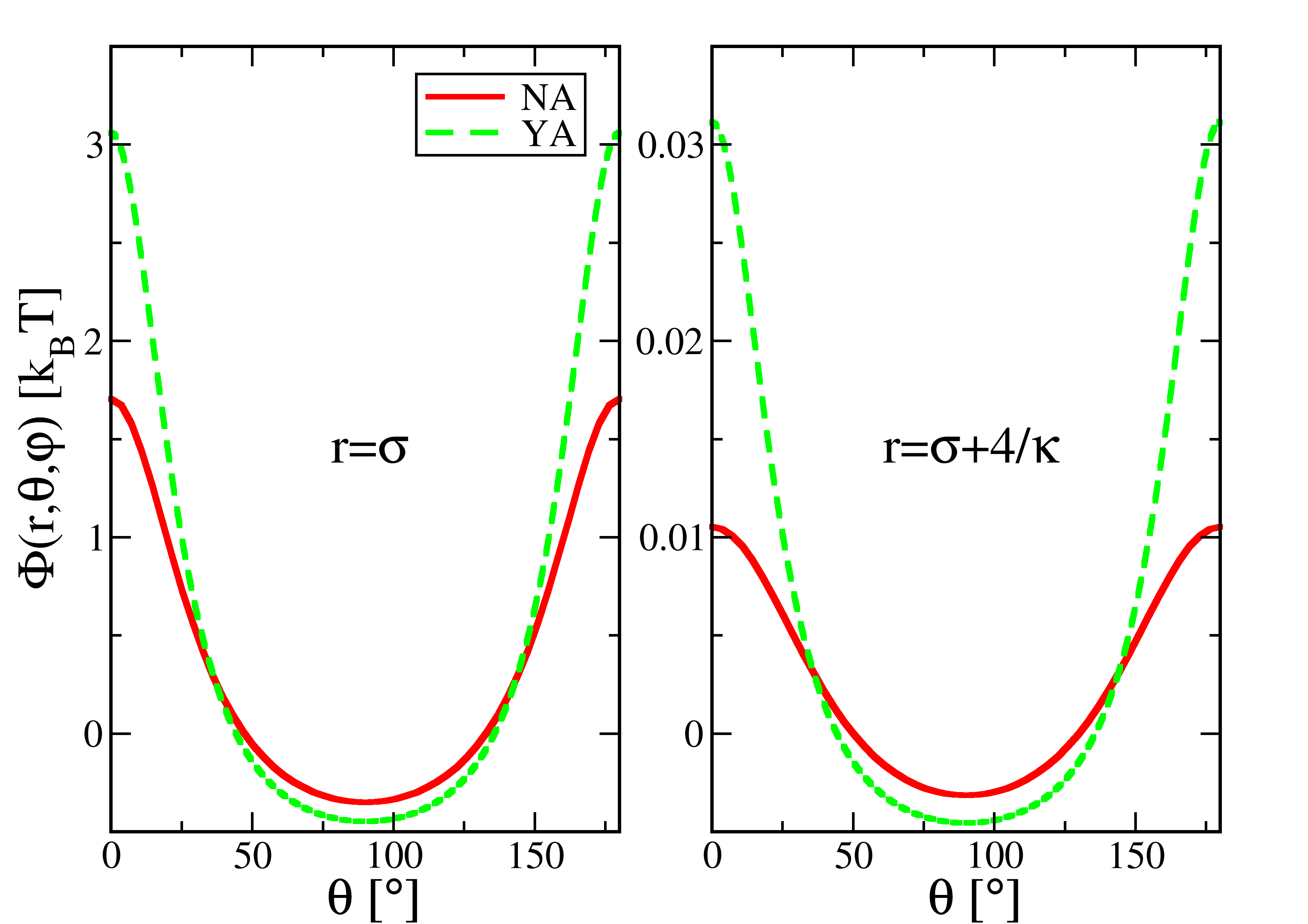}
\caption{Single particle potential $\Phi(r,\theta,\varphi)$ for an IPC with two identical patches; the potential is represented as a function of the angle $\theta$ at a fixed distance:  $r=\sigma$ (left panel) and $r=\sigma+\frac{4}{\kappa}$ (right panel); for two identical patches, the single particle potential does not depend on $\varphi$ due to the azimuthal symmetry.  The following set of parameters is chosen: $Z_{\rm c}=-180$, $Z_{\rm p_1}=Z_{\rm p_2}=Z_{\rm p}=90$, $a_1=a_2=a=0.6\sigma$, $\kappa\sigma=5$ and $\epsilon=80$ (dielectric permittivity of water at room temperature). In both panels,  the continuous red line corresponds to the numerical approach (NA) developed here -- with cut-off $l_{\rm max}=80$ for the number of spherical harmonics --, while the dotted green line reproduces the data obtained via the Yukawa approximation (YA) used in Ref.~\cite{bianchi:ipcfirst}.}
\label{fig:fig1}
\end{figure}

In Figure \ref{fig:fig1} we show the single particle potential $\Phi(r, \theta, \varphi)$ generated by an IPC carrying two identical patches as a function of $\theta$, keeping $r$ fixed to either $r = \sigma$ (left panel) or $r = \sigma + 4/\kappa$ (right panel).  The system parameters used for the comparison are specified in the figure caption.  In these panels we compare results calculated via the analytic route (presented here) and data obtained via the approximate approach put forward in Ref.~\cite{bianchi:ipcfirst}; the two different routes will be labeled henceforward as ``numerical approach'' (NA) and ``Yukawa approximation'' (YA), respectively.

We note that even though the potential reported in Ref.~\cite{bianchi:ipcfirst} can formally be obtained by considering only the lowest-order spherical harmonics, these data cannot be reproduced within the present formalism by simply setting $\ell_{max}=0$. This discrepancy can be traced back to how the cut-off is conducted in the different approaches: in Ref.~\cite{bianchi:ipcfirst} the sum over the
Legendre polynomials is approximated {\it after} the total potential has been calculated analytically; in contrast, in the numerical calculation, the approximation needs to be carried out directly in the basis functions, {\it before} the boundary conditions are imposed. 
%

\subsection{The effective potential between two IPCs}\label{sec:pairpot}

Once the single particle potential is known, the effective interaction energy between two identical colloids can be calculated using the same procedure as put forward in Ref.~\cite{bianchi:ipcfirst}: we first determine the potential energy due to the presence of an IPC in the screened electrostatic field generated by another IPC (see Equations (5) and (6) of Ref.~\cite{bianchi:ipcfirst}) and then we symmetrize it; finally, the total interaction energy for a given particle-particle configuration is obtained as the average value over the two contributions divided by the absolute value of the minimum of the attraction. 

The described procedure is analytically applied in Ref.~\cite{bianchi:ipcfirst} due to the Yukawa approximation operated on the single particle potential; however, in the present contribution we calculate the pair potential via a numerical route, thus retaining higher-order spherical harmonics. 

The radial and angular dependence of the resulting pair potential is investigated in Sec.~\ref{sec:results} on changing the microscopic parameters of the model;  in particular, we focus on the effect of varying either the surface charges, i.e. $Z_{\rm c}$, $Z_{\rm p_1}$, and $Z_{\rm p_2}$, or the patch positions, i.e. $a_1$ and $a_2$, while keeping the screening conditions, imposed by $\kappa\sigma$, constant.

\section{The coarse-grained description}\label{sec:coarse-grained}

We consider the coarse-grained description introduced in Ref.~\cite{bianchi:ipcfirst} for  IPCs characterized by two identical patches and we extend this model to the case of two different patches. The coarse-grained description is designed to reproduce the same symmetries as the underlying microscopic system and is characterized by three independent sets of parameters: the interaction ranges of the different surface regions, their interaction strengths, and their surface extents. 

The model features a hard spherical particle of size $\sigma$ carrying two interaction sites placed in opposite directions at distance $a_1$ and $a_2$ from the particle center; both distances are smaller than $\sigma$ so that the two sites are always located inside the colloid. As a consequence, the corresponding site interaction spheres (with radii $\rho_1$ and $\rho_2$) extend only partially outside the hard core particle, defining in this way the polar patches; the respective surface extensions of the patches are characterized by the half opening angles $\gamma_1$ and $\gamma_2$. 

Since the characteristic interaction distances in the microscopic system are determined by the electrostatic screening of the surrounding solvent, all entities of the colloid are assumed to have the same interaction range, $\delta$, irrespective of the surface regions involved in the interaction.  The interaction sphere of the bare colloid has thus radius $\sigma+\delta/2$. 

By construction, the following relations between the {\it geometric} parameters of the model hold
\begin{eqnarray}
\frac{\delta}{2}&=&a_i+\rho_i-\sigma \\
\cos\gamma_i&=&\frac{\sigma^2+a_i^2-\rho_i^2}{2\sigma a_i}
\end{eqnarray}
where $i=1, 2$; these relations are the generalizations of Equations~(10) and (11) in Ref.~\cite{bianchi:ipcfirst}, where only the special case $a_1\equiv a_2$ and $\rho_1 \equiv \rho_2$ was studied. By virtue of these constraints, the model is characterized only by two sets of geometric parameters: the particle interaction range $\delta$ and the patch sizes $\gamma_1$ and $\gamma_2$. Indeed, once the independent parameters of the model, $a_{1,2}$ and $\rho_{1,2}$, are defined, the physical parameters, $\delta$ and $\gamma_{1,2}$, are also fixed.

In contrast, the {\it energy} parameters of the model are related to the charges involved in the interactions. While the center of the particle carries a charge $Z_{\rm c}$, the two sites carry charges $Z_{\rm p_1}$ and $Z_{\rm p_2}$, respectively. These charges are responsible for the ratio between the attractive and repulsive contributions to the pair energy associated to the different (patch/patch, patch/bare and bare/bare) interactions. 

The specific form of the pair potential is based on the postulate that each of the aforementioned contributions can be factorized into an energy strength and a geometrical weight factor, the latter one being given by the distance dependent overlap volume of the involved interaction spheres. More specifically, beyond the hard core repulsion, the pair potential between two IPCs at distance $r$ is given by~\cite{bianchi:ipcfirst} (omitting the arguments for brevity)
\begin{equation}\label{eq:U}
U  = 
\left\{
\begin{array}{rl}
\frac{3}{4\pi\sigma^3}\sum_{ij} u_{ij}w_{ij} &{\hspace{1em}\rm if\hspace{1em}} 2\sigma<r<2\sigma+\delta \\
0                          &{\hspace{1em}\rm if\hspace{1em}} r \ge 2\sigma+\delta,
\end{array}
\right. ;
\end{equation}
here $i$ ($j$) specifies either a site or the center of the first (second) IPC, $w_{ij}$ is the overlap volume of the corresponding interaction spheres, and $u_{ij}$ is the energy strength of the $ij$ interaction. We note that, while the $u_{ij}$ are constants fixed by the mapping procedure, the $w_{ij}$ -- as well as the potential $U$ -- depend on both the inter-particle distance and the relative orientation of the two IPCs. Although complex functions of the particle-particle distance and orientation, the $w_{ij}$ can be written as simple functions of the site-site, site-center and center-center distances, as reported in Ref.~\cite{bianchi:ipcfirst}.

Both the length and the energy scales involved in the coarse-grained model are imposed by the mapping schemes outlined in Ref.~\cite{bianchi:ipcfirst}. For what concerns the length scales, we assume that the coarse-grained interaction range is proportional to the Debye screening length according to the following relation: $\kappa\delta=n$, where $\kappa$ is determined by the screening conditions $\kappa\sigma=m$; thus $\delta=\frac{n}{m}\sigma$, where $m$ and $n$ are not necessarily integer numbers. For a given $\delta$, the angular patch extents $\gamma_{1,2}$ are defined by the choice of either $a_{1,2}$ or $\rho_{1,2}$. The overlap volumes $w_{ij}$  for a given configuration are fixed by the choice of the geometric parameters. On the other hand, the energy scales are imposed by the mapping scheme referred to as ``max'' in Ref.~\cite{bianchi:ipcfirst}: the energy strengths $u_{ij}$ are fixed by considering characteristic reference configurations of two IPCs at contact and then imposing that the contact values of the Debye-H\"{u}ckel and the coarse-grained potentials are equal. For a given set of geometric parameters, the overall particle charge $Z_{\rm tot}=Z_{\rm c}+Z_{\rm p_1}+Z_{\rm p_2}$ determines the set of $u_{ij}$ values for the chosen reference configurations.

For IPCs with two symmetric patches the three configurations depicted on the right hand side of Figure~\ref{fig:fig2}, namely the polar-polar (PP), the equatorial-polar (EP) and the equatorial-equatorial (EE) configuration, are enough to fully describe the inter-particle potential. In this case, the pair interaction energy is normalized with the value corresponding to the EP configuration~\cite{bianchi:2d2013,bianchi:2d2014}. For the full description of IPCs with asymmetric patches (either in size or in charge), six characteristic particle configurations are needed: in particular, in the PP and EP configurations the existence of two different patches must be taken into account. In the following we denote differently charged or sized patches as patch 1 and 2, and we thus distinguish six characteristic configurations, i.e. PP(11), PP(12), PP(22), EP(1), EP(2), and EE. In this case, the interaction energies are normalized by the most negative value occurring (i.e. the most attractive interaction), corresponding either to the EP(1) or the EP(2) configuration.

\section{Results}\label{sec:results}

In the following we evaluate the effective potentials between different pairs of two-patch IPCs in characteristic particle-particle configurations. More specifically, we investigate in subsection~\ref{sec:DH} the radial and angular dependence of the Debye-H{\"u}ckel potential, while we discuss the coarse-grained counterpart in subsection~\ref{sec:CG}.

\subsection{The Debye-H{\"u}ckel potential}\label{sec:DH}

In the following we first consider the case of IPCs with symmetric patches and compare the Debye-H{\"u}ckel potentials obtained with the NA developed here to the results obtained within the YA proposed in Ref.~\cite{bianchi:ipcfirst} (see~\ref{sec:symm}); we then discuss the NA-based potentials for IPCs with asymmetric patches, either in charge (see subsection~\ref{sec:asymm-c}) or in size (see subsection~\ref{sec:asymm-s}).

\subsubsection{Symmetric patches}\label{sec:symm}

\begin{figure}[!h]
\includegraphics[width=0.5\textwidth]{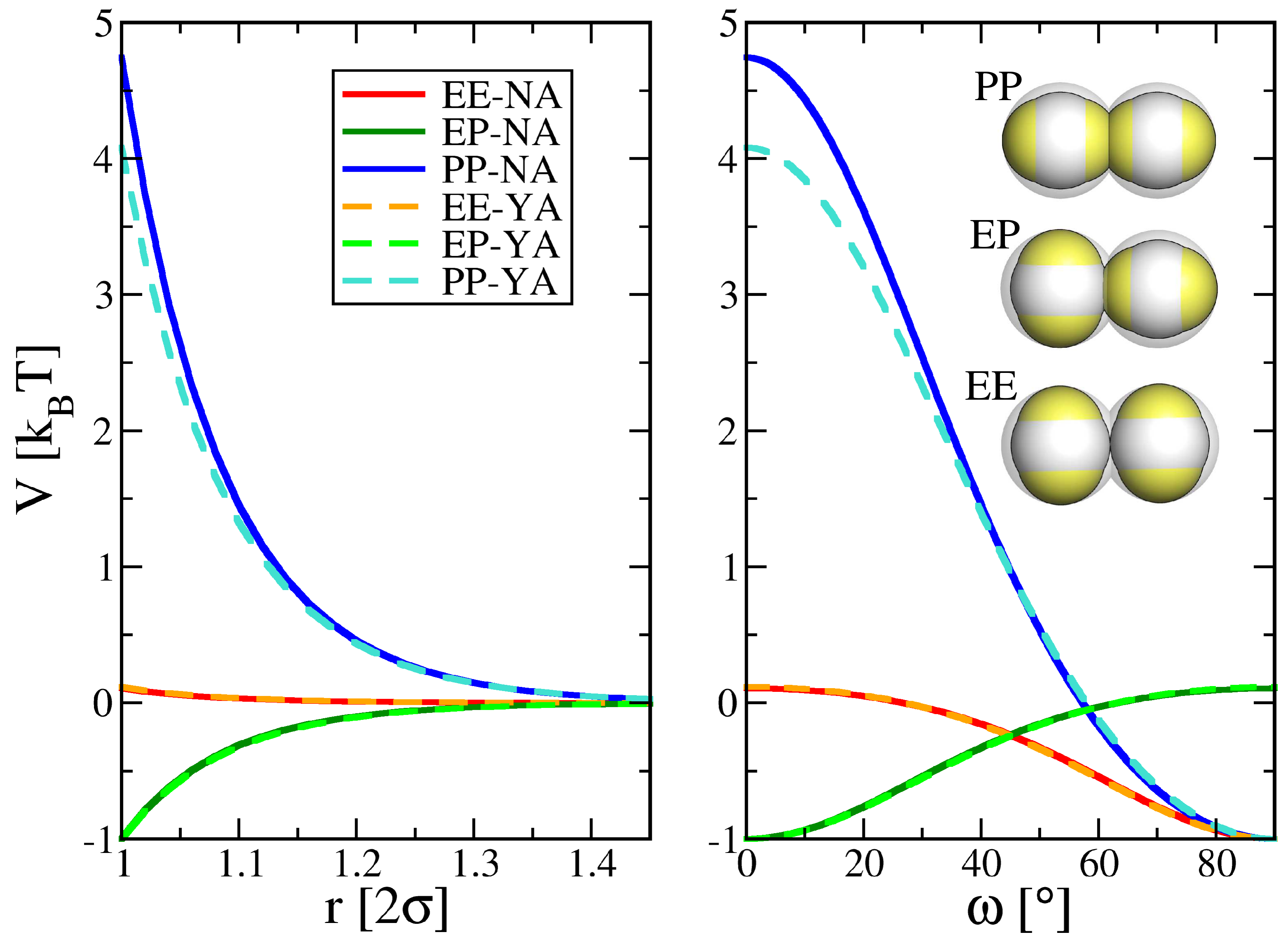}
\caption{Radial (left) and angular (right) dependence of the interaction energy $V$ (in units of $k_{\rm B}T$) of two overall neutral IPCs with two identical patches. The specific choice of the parameter set guarantees that data obtained within the YA correspond to the model referred to as 45n in Ref.~\cite{bianchi:2d2013}: $Z_{\rm c}=-180$, $Z_{\rm p_1}=Z_{\rm p_2}=Z_{\rm p}=90$, $a_1=a_2=a=0.44\sigma$, $\kappa\sigma=5$, and $\epsilon=80$ (dielectric permittivity of water at room temperature). Results obtained within the NA are represented by continuous lines, while YA-based data are depicted by dotted lines. In the left panel, the potentials are shown as functions of the inter-particle distance $r$ between two IPCs in three different mutual orientations (as labeled): the polar-polar (PP), the equatorial-polar (EP), and the equatorial-equatorial (EE) configurations, depicted schematically in the right panel; in the right panel, the potentials are represented for two IPCs at contact, i.e. at a distance $r=2\sigma$, as functions of the rotation angle $\omega$ around an axis perpendicular to the plane of the figure: after a rotation of $90^{\degree}$ one particle-particle configuration transforms into one of the other particle arrangements.}
\label{fig:fig2}
\end{figure}

We first consider overall neutral IPCs with two identical patches, i.e., we set $Z_{\rm tot} = 0$, $Z_{\rm p}=Z_{\rm p_1}=Z_{\rm p_2}$, and $a=a_1=a_2$. For this case we compare the radial and angular dependence of the pair potentials obtained within the NA developed in this contribution and within the YA put forward in Ref.~\cite{bianchi:ipcfirst}. To this end we consider a selected set of parameters: in particular, we fix the screening conditions $\kappa \sigma$, the charges $Z_{\rm c}$ and $Z_{\rm p}$, and the patch-center distance $a$ such that we recover the same set of model parameters referred to as 45n in Ref.~\cite{bianchi:2d2013}; this label identifies the selected model by the patch size of the related coarse-grained description, i.e., $\gamma_1=\gamma_2=45^{\degree}$, and by the overall particle charge, where ``n'' stands for neutral. In the following, we continue to use this labeling scheme. As shown in Fig.~\ref{fig:fig2}, the results obtained within the two approaches are sufficiently close to each other: only in the case of the PP repulsion we observe that it is slightly stronger within the NA; for the chosen set of parameters (specified in the caption of Fig.~\ref{fig:fig2}), the NA contact energy of the PP configuration is by a factor $\approx 1.15$ bigger than the corresponding YA-value. 
 
\begin{figure}[!h]
\includegraphics[width=0.5\textwidth]{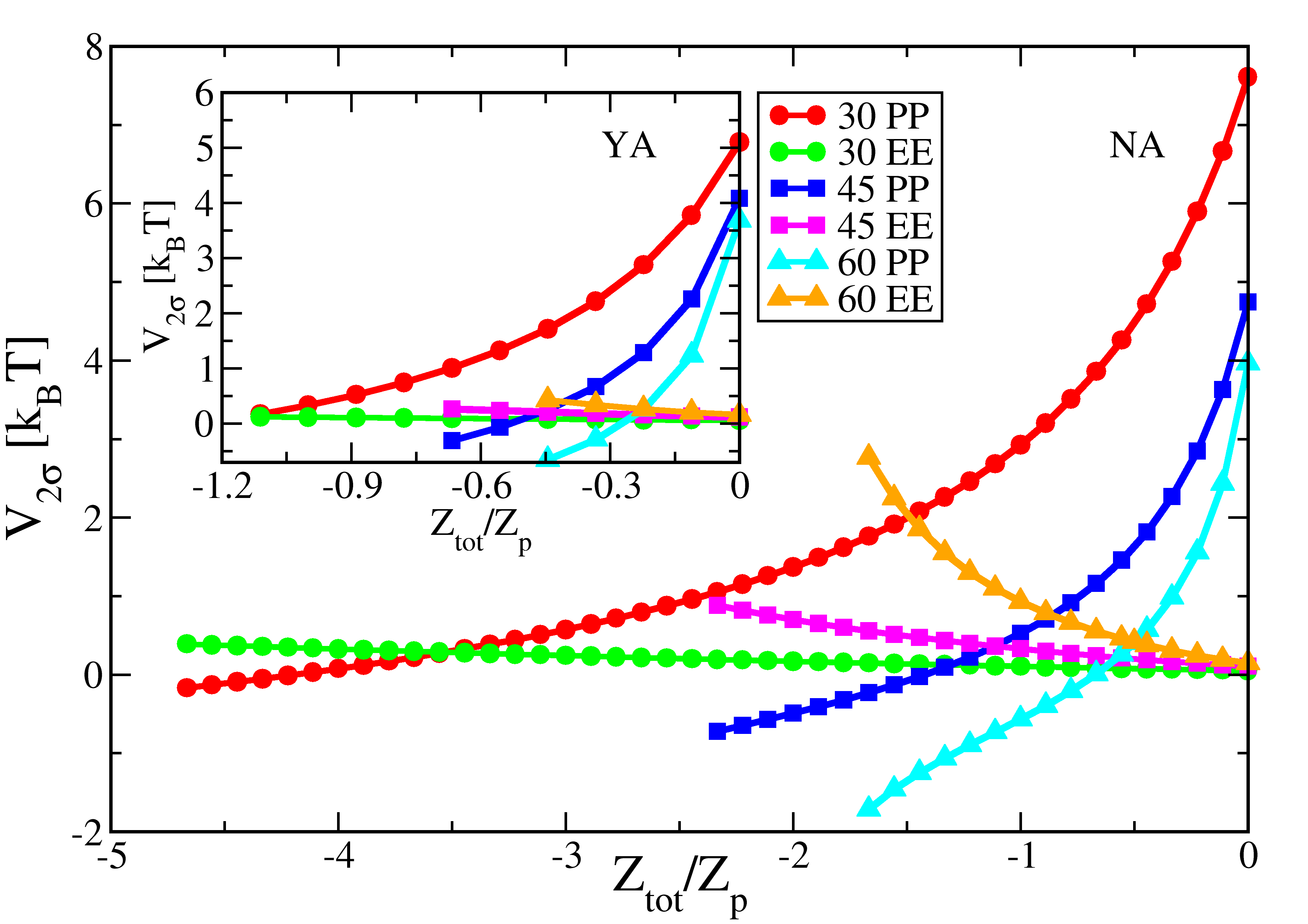}
\caption{Contact energies $V_{2\sigma}$ (in units of $k_{\rm B}T$) of two IPCs at contact in the EE and PP configurations (as labeled) as functions of the charge imbalance $Z_{\rm tot}/Z_{\rm p}$; the contact energy between two IPCs in the EP configuration is by construction equal to -1 and is thus not displayed. The main panel reports the NA-results, while the inset shows the corresponding data obtained within the YA. For fixed screening conditions and particle charges,  three patch-center distances (corresponding to three different patch sizes in the related coarse-grained models) are chosen such that the overall neutral systems studied in Refs.~\cite{bianchi:2d2013,bianchi:2d2014} are recovered. For overall neutral IPCs, the following set of parameters is chosen: $Z_{\rm c}=-180$, $Z_{\rm p_1}=Z_{\rm p_2}=Z_{\rm p}=90$, $a_1=a_2=a=0.64\sigma$ (labeled 30 to specify the coarse-grained patch size $\gamma=30^{\degree}$), $0.44\sigma$ (labeled 45 to highlight the coarse-grained patch size $\gamma=45^{\degree}$), and $0.32\sigma$ (labeled 60 to specify the coarse-grained patch size $\gamma=60^{\degree}$), $\kappa\sigma=5$ and $\epsilon=80$ (dielectric permittivity of water at room temperature); for overall charged IPCs, the patch charges are fixed to $Z_{\rm p}=90$, while the colloid charge is changed from $Z_{\rm c}=-180$ by $Z_{\rm tot}$.}
\label{fig:fig3}
\end{figure}

As the contact energy,  $V_{2\sigma}$, for the different particle-particle configurations characterizes the pair potential, we calculate such a value on varying the charge imbalance between the patches and the bare colloid, starting from the 45n model and increasing the colloid charge $Z_{\rm c}$; the charge imbalance is thus quantified by the overall particle charge $Z_{\rm tot}$. The corresponding data, labeled with 45, are reported in Fig.~\ref{fig:fig3}: in the main panel we show the NA-data, while in the inset YA-results are displayed. Since for the EP configuration  $V_{2\sigma}$  is always equal to unity by construction, only the contact values for the repulsive contributions are reported. Both approaches show the same trend on increasing $Z_{\rm tot}$: while in the overall neutral system the PP repulsion is much stronger than the EE repulsion, in overall charged systems the PP repulsion decreases whereas the EE one increases up to the point where the EE interaction becomes more repulsive than the PP one. The comparison between the NA and the YA reveals that even though this trend is present in both sets of data, the values of $Z_{\rm tot}$ at which the PP and the EE repulsions are equal differ between the two approaches: for the chosen set of parameters (specified in the caption of Fig.~\ref{fig:fig3}), $Z_{\rm tot}/Z_{\rm p} \simeq -1.11$ within the NA, while this value is approximately -0.44 within the YA, indicating that the described trend of the characteristic repulsions is more pronounced in the latter potential description. 

\begin{figure}[!h]
\includegraphics[width=0.5\textwidth]{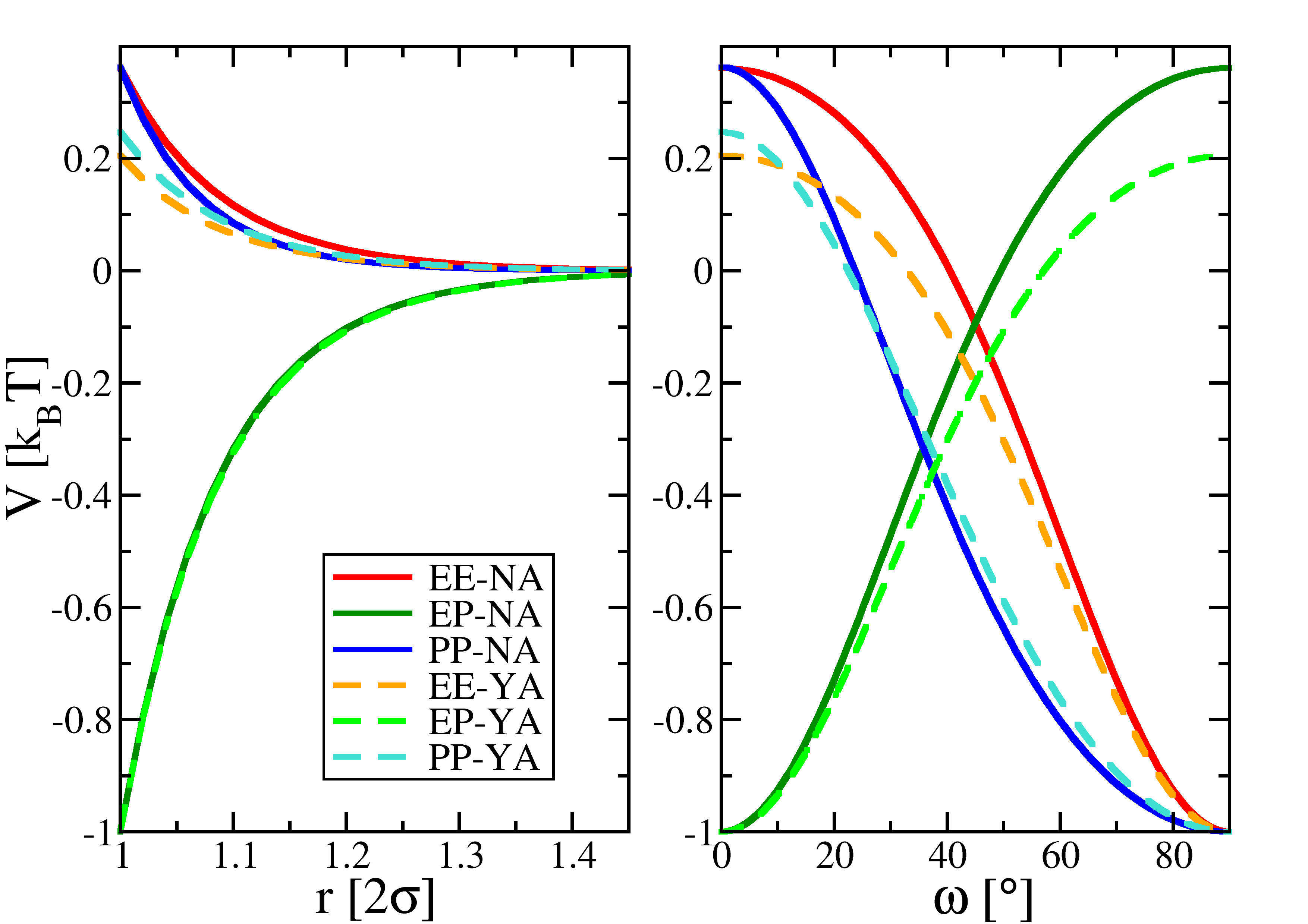}
\caption{Radial (left) and angular (right) dependence of the interaction energy $V$ (in units of $k_{\rm B}T$) between two overall charged IPCs with two identical patches. The specific choice of the parameter set guarantees that data obtained within the YA correspond to the model referred to as 45c in Ref.~\cite{bianchi:2d2013}: $Z_{\rm c}=-280$ (for the NA) or $-220$ (for the YA), $Z_{\rm p_1}=Z_{\rm p_2}=Z_{\rm p}=90$, $a_1=a_2=a=0.44\sigma$, $\kappa\sigma=5$, and $\epsilon=80$ (dielectric permittivity of water at room temperature). Results for the NA are represented by continuous lines, while data obtained with the YA approach are depicted by dotted lines. In the left panel, the potentials are reported as functions of the inter-particle distance $r$ between two IPCs in the PP, EP, and EE configurations (as labeled); in the right panel, the potentials are represented for two IPCs at contact, i.e. at a distance $r=2\sigma$, as functions of the rotation angle $\omega$ around an axis perpendicular to the plane of the figure: after a rotation of $90^{\degree}$ one particle-particle configuration transforms into one of the other particle arrangements.}
\label{fig:fig4}
\end{figure}

The radial and angular dependence of the pair potential between charged IPCs characterized by almost identical EE and PP contact energies is reported in Fig.~\ref{fig:fig4}: for the chosen set of parameters (specified in the caption of Fig.~\ref{fig:fig4}) we fix in our NA and YA calculations $Z_{\rm tot}/Z_{\rm p}=-1.11$ and $Z_{\rm tot}/Z_{\rm p}=-0.44$, respectively; the latter system corresponds to the set of model parameters that is referred to as 45c in Ref.~\cite{bianchi:2d2013}; again, such a label identifies the selected model via the patch size of the related coarse-grained description, i.e., $\gamma_1=\gamma_2=45^{\degree}$, and by the overall particle charge, where ``c'' stays for charged. As in the neutral case, the trends in the radial and angular dependencies of the potentials are the same for both approaches, the NA providing data that are only slightly more repulsive than the YA-based results (for both the EE and the PP configurations).

Finally, we studied the behavior of the contact energy as a function of the charge imbalance for other IPC models, specified by a different choice of the parameter $a$. We note that at fixed interaction range, the choice of $a$ uniquely determines the patch size $\gamma$ of the related coarse-grained model. In Fig.~\ref{fig:fig3} we report $V_{2\sigma}$-values for IPCs that are labeled by 30 (with a patch-center distance bigger than the one of the 45 model and such that $\gamma=30^{\degree}$) and 60 (with a patch-center distance smaller than the one of the 45 model and such that $\gamma=60^{\degree}$), assuming EE and PP configurations. The choice of the parameters (specified in the caption of Fig.~\ref{fig:fig3}) is such that for $Z_{\rm tot} = 0$ the YA reproduces the interactions of the overall neutral models studied in Ref.~\cite{bianchi:2d2013} (referred to as 30n) and in Ref.~\cite{bianchi:2d2014} (referred to as 60n); accordingly, their overall charged counterparts investigated in Refs.~\cite{bianchi:2d2013,bianchi:2d2014} are reproduced by the YA potentials for $Z_{\rm tot}/Z_{\rm p}=-1.11$ (model referred to as 30c) and $-0.22$ (model referred to as 60c). Despite a scaling factor, the NA and YA provide very similar trends for the repulsive contact energies in the PP and the EE configurations as functions of the charge imbalance. Thus, we can speculate that also for models 30 and 60 the full radial and angular dependencies of the corresponding potentials will not show significant differences between the two approaches, as shown for model 45. We can conclude that both the NA and the YA provide consistent results for the description of the inter-particle interaction between IPCs.

\subsubsection{Asymmetric patches in charge}\label{sec:asymm-c}
\begin{figure}[!h]
\includegraphics[width=0.5\textwidth]{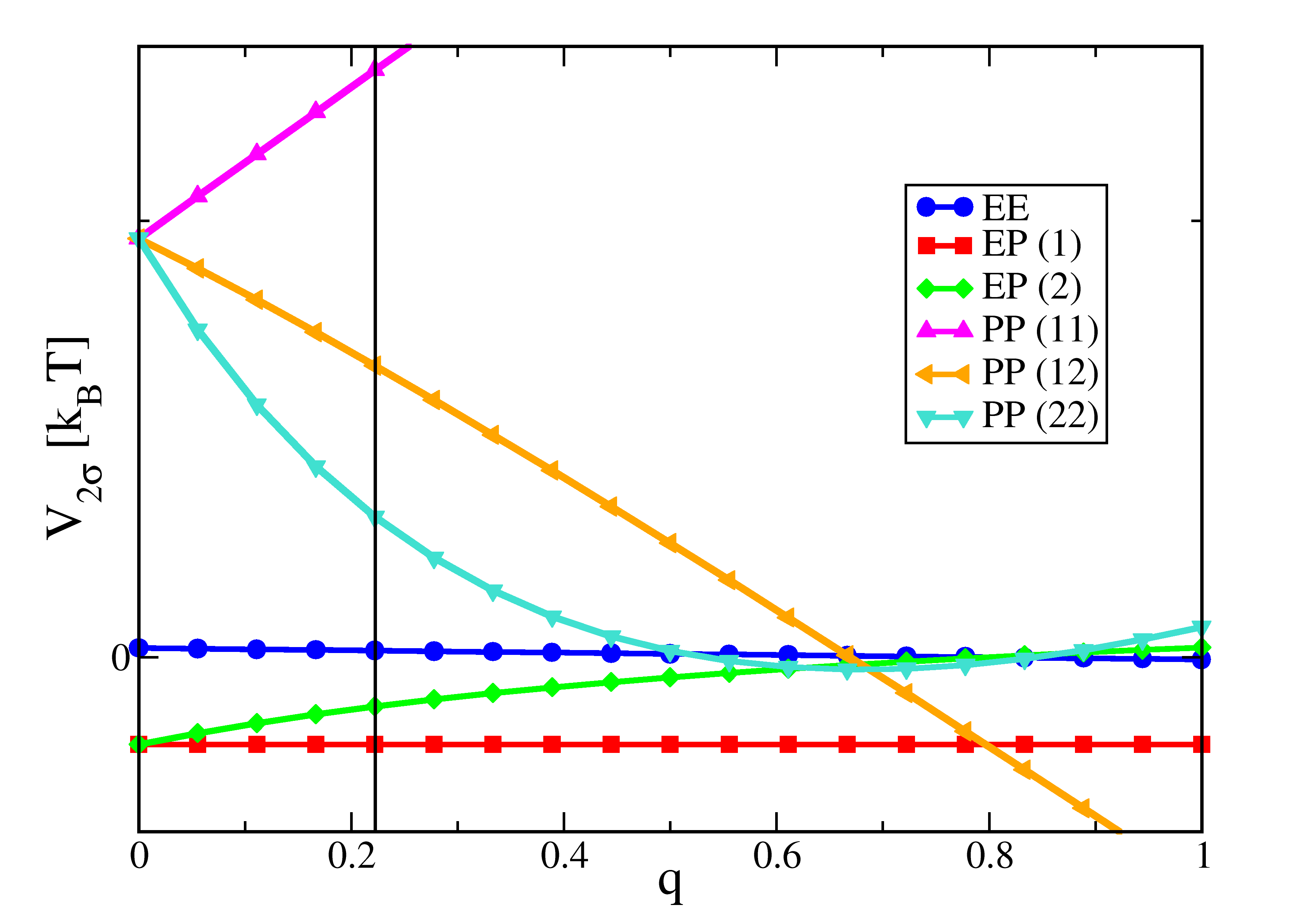}
\caption{Contact energies $V_{2\sigma}$ (in units of $k_{\rm B}T$) between two overall neutral IPCs in six different mutual orientations (as labeled) as functions of the charge asymmetry parameter $q$ (defined in the text). When $q=0$, the model corresponds to the 45n model whose parameters are given in the caption of Fig.~\ref{fig:fig2}. The models with differently charged patches ($q > 0$) are obtained in the following way: the central charge is kept constant to $Z_{\rm c}=-180$ and the two patch charges, initially set to the same value $Z_{\rm p_1}=Z_{\rm p_2}=Z_{\rm p}=90$, are either increased or decreased by a factor $q$, i.e. $Z_{\rm p_1}=Z_{\rm p}(1+q)$ and $Z_{\rm p_2}=Z_{\rm p}(1-q)$; $Z_{\rm p}$ sets the charge unit. The vertical black line highlights the value of $q=0.22$ for which the radial and angular dependence of the pair potentials is shown in Figure~\ref{fig:fig6}.}
\label{fig:fig5}
\end{figure}

We now consider IPCs with two polar patches of equal size but different charge. In particular, we focus on the 45n model investigated in the previous subsection and we change the charges of the two patches, while keeping the central charge of the colloid fixed. In order to maintain a vanishing overall charge of the particle, the charge of one patch (say $Z_{{\rm p}_1}$) is increased from $Z_{\rm p}$  to $Z_{\rm p} (1 + q)$, while the charge of the other patch (say $Z_{{\rm p}_2}$) is decreased from $Z_{\rm p}$ to $Z_{\rm p} (1 - q)$. In Fig.~\ref{fig:fig5} we report the resulting contact energy values as functions of $q$ for the six characteristic particle-particle configurations specified in Sec.~\ref{sec:coarse-grained}. As expected the three PP energies, as well as the two EP energies, are identical for $q=0$ and they deviate from each other as soon as $q$ increases. We note that since the $V_{2\sigma}$-value for IPCs in the EP(1) configuration is used to normalize the pair potential, the corresponding curve in the ($V_{2\sigma}, q)$-plane is of course a horizontal line; on the other hand, the EP(2) value for $V_{2 \sigma}$ increases from -1 to slightly positive values as the charge of the second patch decreases, i.e. on increasing $q$. We also note that the contact value of the EE repulsive interaction is almost constant -- it changes from slightly positive to slightly negative values -- as the patch-free zones are unaffected by the change of the patch charges, at least as long as the IPC is overall neutral. Finally, as  $Z_{{\rm p}_2}$ increases and  $Z_{{\rm p}_1}$ decreases, the PP(11) and PP(22) contact values rapidly increase and decrease, respectively; in contrast, the PP(12) contact value, being given by the overlap of one decreasing and one increasing patch charge, features a much slower decrease. In Fig.~\ref{fig:fig5} there is a narrow $q$-range where the curves of the PP(12), PP(22), EE and EP(2) contact energies intersect (i.e., at $q^* \approx 0.66$ for the particular set of parameters chosen). Beyond this point the PP(22) energy becomes attractive although the patches still have the same sign. This value depends on the system parameters: at fixed interaction range, the $q$-value at which all the interactions -- except the reference EP(1) one -- have very similar contact energies increases as the patch size decreases. Nonetheless, since at this point the charge asymmetry is already quite large, the potential description must be considered unreliable. 

\begin{figure}[!h]
\includegraphics[width=0.5\textwidth]{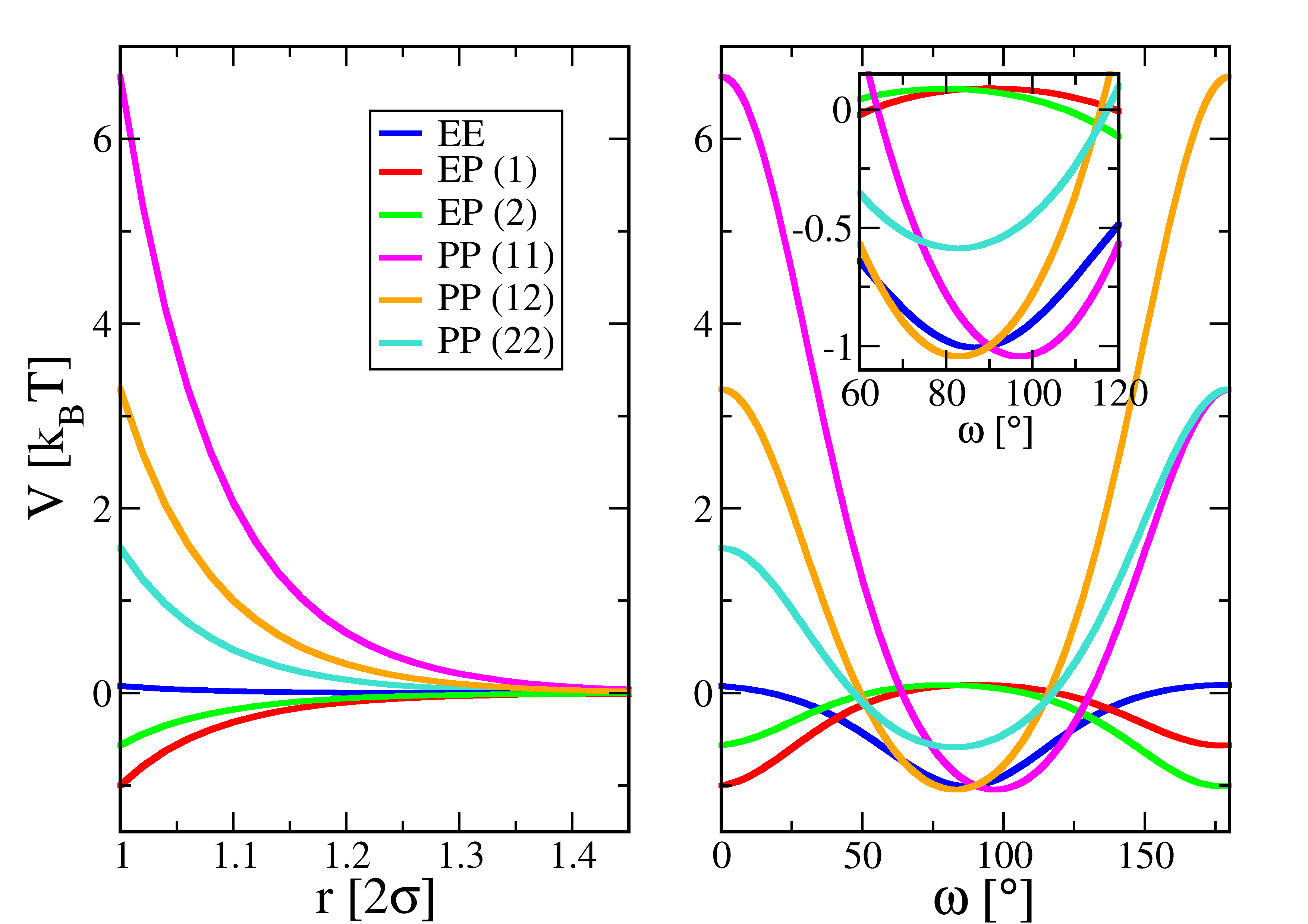}
\caption{Radial (left) and angular (right) dependence of the interaction energy $V$ (in units of $k_{\rm B}T$) between two overall neutral IPCs at contact carrying two differently charged patches. The asymmetry between the two patch charges is specified by $q=0.22$ (highlighted by the black vertical line in Fig.~\ref{fig:fig5}). In the left panel, the potentials are reported as functions of the inter-particle distance $r$ between two IPCs in six characteristic particle-particle configurations (as labeled); in the right panel, the potentials are represented for two IPCs at contact, i.e. at a distance $r=2\sigma$, as functions of the rotation angle $\omega$ around an axis perpendicular to the plane of the figure: after a rotation of $180^{\degree}$ one particle-particle configuration transforms into another. In the inset of the right panel a magnified view of the potentials around $\omega=90^{\degree}$ is shown.}
\label{fig:fig6}
\end{figure}

In Fig.~\ref{fig:fig6} we show the full radial and angular dependencies of the inter-particle potential for a IPC model with asymmetrically charged polar patches, specified by $q=0.22$. An interesting effect of the asymmetry is that the positions of the various minima and maxima in the angular dependence of the potentials are slightly shifted to larger or smaller angles with respect to the symmetric case. A detailed view of the potentials for $\omega \simeq 90^{\degree}$ in the right panel of Fig.~\ref{fig:fig6} shows that, for example, the minimum of the EP(1) attraction is not located at $\omega=90^{\degree}$ but rather at $83^{\degree}$ for the chosen set of parameters. The optimal particle-particle configuration is thus such that the less charged patch of one IPC is closer to the more charged patch of the other IPC: for the specific set of parameters, the former is shifted by $7^{\degree}$ closer to the latter with respect to the perfect EP configuration; as a consequence the distance between the centers of the more charged patches on the two different IPCs is bigger than their distance in the perfect EP configuration. We observe that, since different shifting angles are found for different fashions of rotational freedom, the optimal relative orientation may not be covered by the rotations studied here. This asymmetry effect is more pronounced as the charge asymmetry parameter $q$ increases, so that only a small $q$-range can be studied with a reasonable accuracy within our framework.

Investigations on models 30n and 60n (not shown here) have highlighted that for a given $q$-value, the shift of the minima and maxima in $\omega$ is more pronounced as the patch size increases.

\subsubsection{Asymmetric patches in size}\label{sec:asymm-s}
Finally, we consider IPCs with two polar patches of equal charge but of different size. We start again from the 45n model and now change both the interaction range of one polar site, e.g. the site corresponding to patch 2, and its distance from the particle center in such a way that the quantity $a_2+\rho_2$ remains constant. This procedure guarantees that in the corresponding coarse-grained system, the interaction range of patch 2 is the same as the interaction range of the whole particle, i.e. $\delta$ is fixed. For a specific interaction range, there is thus only one choice for $a_2$ for a given $\rho_2$, and vice versa; as a consequence there is only one the patch size $\gamma_2$ associated to the choice of $\rho_2$ and $a_2$ or, equivalently, of $\rho_2$ and $\delta$. 

\begin{figure}[!h]
\includegraphics[width=0.5\textwidth]{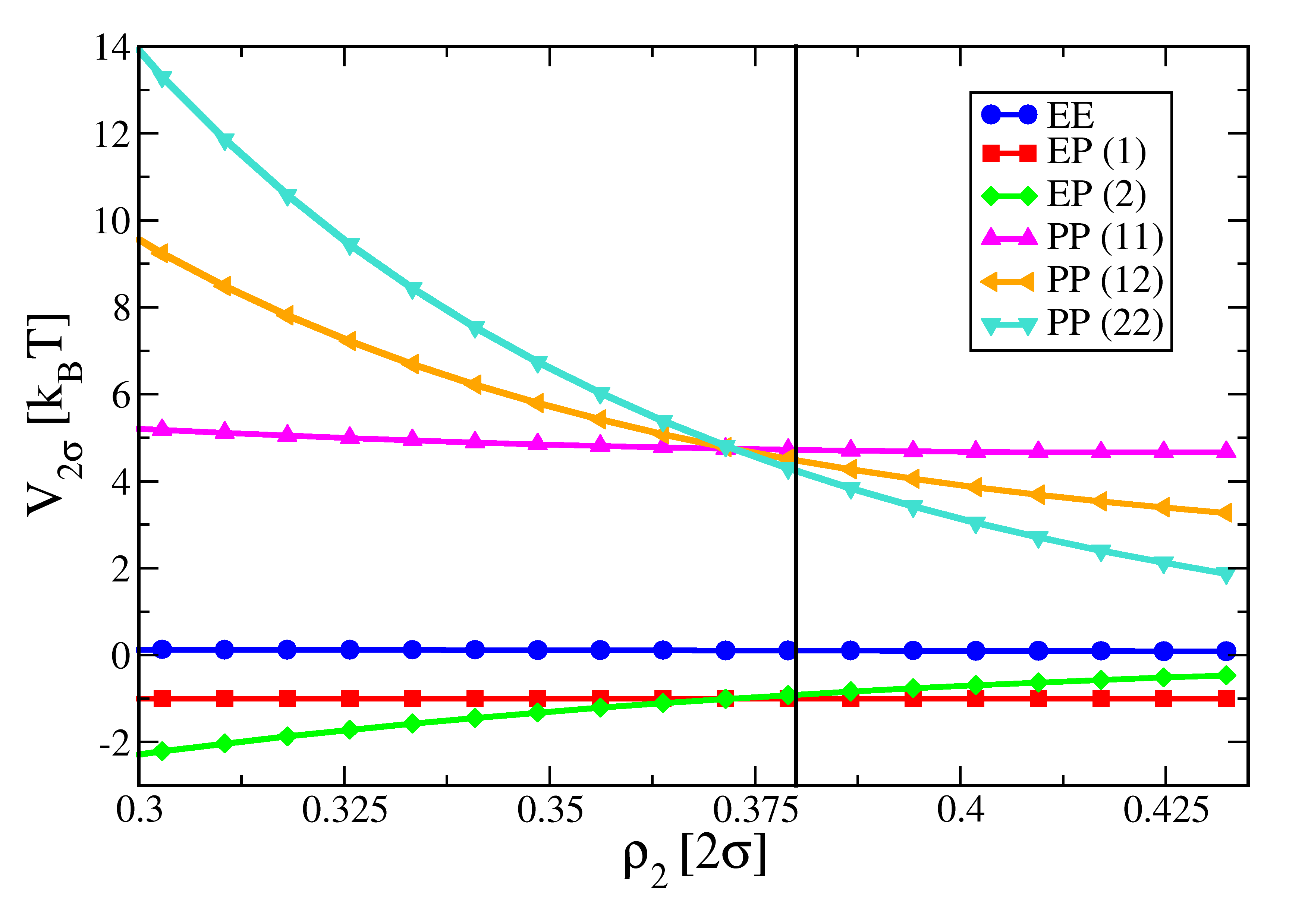}
\caption{Contact energies $V_{2\sigma}$ (in units of $k_{\rm B}T$) between two overall neutral IPCs in six different mutual orientations (as labeled) as a function of the radius of one patch, $\rho_2$, while the radius of the other patch is fixed to $\rho_1=0.76\sigma$. The charges of the patches are equal. The black vertical line marks the set of parameters that corresponds to the 45n model; left to the line the size of patch 2 decreases, while right to the line its extent increases.}
\label{fig:fig7}
\end{figure}

In Fig.~\ref{fig:fig7} we report the behavior of the contact energy values $V_{2 \sigma}$ of the selected IPCs as functions of $\rho_2$. The vertical line in the figure marks the size of patch 1 as well as the size of patch 2 in the 45n model: when $\rho_1=\rho_2$ the three different PP repulsions coincide, while both EP attractions are equal to -1. When $\rho_2$ decreases (i.e. when $\gamma_2$  decreases), both repulsions PP(22) and PP(12) increase, the first one faster than the latter one: when $\rho_2$ is decreased by about 20\%, the PP(22) repulsion has increased its value by a factor of three, while the PP(12) repulsion has only doubled. In both cases the increase of the repulsion is due to the increase of charge per area related to the decrease of patch size for a fixed patch charge. We note that also that the corresponding PP(11) value is slightly affected by a change in the size of patch 2, but its increase amounts to less than 10\% of its initial value; thus it can be considered to be essentially constant. The same consideration applies for the EE repulsion, which changes with the increase of the size of patch 2 by less than 20\%. Since the surface charge increases on decreasing the patch size, also the EP(2) attraction increases: it has more than doubled when $\rho_2$ is decreased by about 20\%; of course, the EP(1) attraction is constant by construction. The opposite trends are observed when $\rho_2$ increases (i.e. when $\gamma_2$  increases); however, the changes in the contact energy values are in this case less pronounced.

\subsection{The coarse-grained potential}\label{sec:CG}

\begin{figure}[h]
\centering
\includegraphics[width=0.5\textwidth]{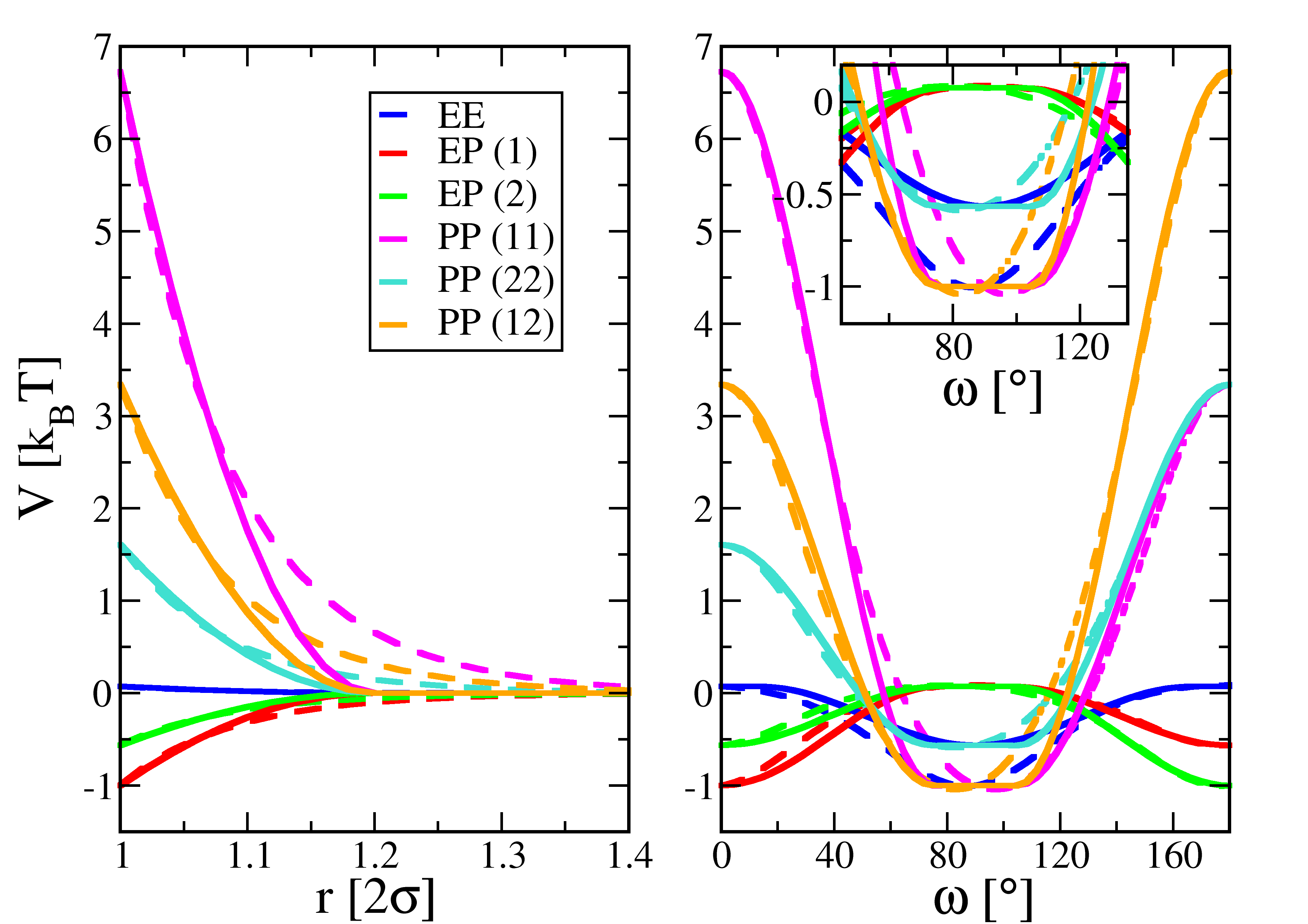}
\caption{Theoretical and coarse-grained pair potentials $V$ (in units of $k_{\rm B}T$) for overall neutral IPCs with patches of size $45^{\degree}$ and charge asymmetry $q=0.22$. Results for the Debye-H{\"u}ckel potential are represented by dotted lines, while those for the coarse-grained interaction (for $\kappa \delta = 2$) are depicted by continuous lines. In the left panel, the potentials are reported as functions of the inter-particle distance $r$ between two IPCs in six different mutual orientations (as labeled); in the right panel, the potentials are represented for two IPCs at contact, i.e. at a distance $r=2\sigma$, as functions of the rotation angle $\omega$ around an axis perpendicular to the plane of the figure: after a rotation of $180^{\degree}$ one particle-particle configuration transforms into another.  In the inset of the right panel a magnified view of the potentials around $\omega=90^{\degree}$ is shown.}
\label{fig:fig8}
\end{figure}

We finally compare the potentials obtained via the Debye-H{\"u}ckel approach and via the coarse-graining procedure for the 45n system with a charge asymmetry parameter $q=0.22$; data are reported in Fig.~\ref{fig:fig8} and show a good quantitative agreement between the Debye-H{\"u}ckel description and the coarse-grained approach. The radial dependence of the potentials is displayed in the left panel of the figure, while their angular dependence is shown in the right panel of the same figure. By definition, at $r=2\sigma$ and $\omega=0^{\degree}$ (i.e. when the two IPCS form the reference configurations) the coarse-grained data coincide with the results obtained in the analytical calculations since the contact energy values are fixed by the mapping procedure. For larger distances (left panel) or angles (right panel) the coarse-grained data follow the analytical ones quite well; however, it must be noted that the angular dependence of the coarse-grained potential around $\omega=90^{\degree}$ shows small differences with respect to the Debye-H{\"u}ckel data. Around the minima and maxima obtained via the analytical approach, the coarse-grained potential forms plateaus that are due to the saturation of the overlap volume between the interaction spheres. The effects described in subsection~\ref{sec:asymm-c} are thus not reproduced by the coarse-grained model: while in the analytical calculations there is an optimal bonding angle, in the coarse-grained description there is a range of angles around the optimal one. As long as this range is reasonably narrow, Monte Carlo simulations based on the coarse-grained potential will give accurate results at finite temperatures. Indeed, when studying ensembles of IPCs in simulations under these conditions small deviations from the analytical pair energy will affect the system behavior only to a minor extent, due to the role played by entropy.

\section{Conclusion}\label{sec:conclusion}

In the present contribution, we have proposed a nearly analytic method to calculate the effective interaction potential between inverse patchy colloids (IPCs) with rich surface patterns. Our description follows the Debye-H\"uckel approach to determine the potential generated by one single colloid dispersed in a liquid solvent: the potentials inside and outside the colloid are calculated separately as expansions in terms of spherical harmonics and are then linked together by electrostatic boundary conditions; the resulting set of linear equations for the yet undetermined expansion coefficients is solved numerically. Once the single particle potential is known, the effective interaction energy between two IPCs is determined as in Ref.~\cite{bianchi:ipcfirst}. The main advantage of not solving the equations analytically is that they can be kept as general as possible so that they can be easily applied to a wide range of different IPCs. The Debye-H\"uckel potential is subsequently used to provide a suitably designed coarse-grained model with parameters that are directly related to the physical quantities of the underlying microscopic system, such as the screening conditions, the surface charges and the extent of the different surface regions on the colloids. The coarse-grained description is advantageous in many body simulations because the model pair energy between two particles is the sum over simple products of geometric (orientation- and distance-dependent) factors and energy factors, associated to the different ways of interaction that characterize the IPC. 

In this contribution, the pair potentials for neutral as well as overall charged colloids with two symmetric or asymmetric patches in size or/and charge are calculated. A comparison between the characteristic energies of the Debye-H\"uckel description and of the coarse grained model shows good agreement. In an effort to demonstrate the versatility of our approach, the framework is also applied to the case of three possibly different patches. 

The proposed method will allow future studies on the self-assembly of heterogeneously charged colloidal systems, that are expected to be observed in experiments.
 
\bibliographystyle{unsrt}
\bibliography{biblio} 

\begin{thebibliography}{10}

\bibitem{kretzschmar:review}
A.~B. Pawar and I.~Kretzschmar.
\newblock {\em Macromol.~Rapid~Commun.}, 31:150, 2010.

\bibitem{bianchi:review}
E.~Bianchi, R.~Blaak, and C.~N. Likos.
\newblock {\em Phys.~Chem.~Chem.~Phys.}, 13:6397, 2011.

\bibitem{Daniel_ACS_2010}
M.-C. Daniel, I.~B. Tsvetkova, Z.~T. Quinkert, A.~Murali, M.~De, V.~M. Rotello,
  C.~{Cheng Kao}, and B.~Dragnea.
\newblock {\em ACS Nano}, 4:3853, 2010.

\bibitem{Zhang_2008}
F.~Zhang, M.~W.~A. Skoda, R.~M.~J. Jacobs, S.~Zorn, R.~A. Martin, C.~M. Martin,
  G.~F. Clark, S.~Weggler, A.~Hildebrandt, O.~Kohlbacher, and F.~Schreiber.
\newblock {\em Phys. Rev. Lett.}, 101:148101, 2008.

\bibitem{Zhang_2012}
F.~Zhang, R.~Roth, M.~Wolf, F.~Roosen-Runge, M.~W.~A. Skoda, R.~M.~J. Jacobs,
  M.~Stzucki, and F.~Schreiber.
\newblock {\em Soft Matter}, 8:1313, 2012.

\bibitem{Christian_NatMat_2009}
D.~A. Christian, A.~Tian, W.~G. Ellenbroek, I.~Levental, K.~Rajagopal, P.~A.
  Janmey, A.~J. Liu, T.~Baumgart, and D.~E. Discher.
\newblock {\em Nat. Mat.}, 8:843, 2009.

\bibitem{Juhl_2006}
S.~B. Juhl, E.~P. Chan, Y.-H. Ha, M.~Maldovan, J.~Brunton, V.~Ward, T.~Dokland,
  J.~Kalmakoff, B.~Farmer, E.~L. Thomas, and R.~A. Vaia.
\newblock {\em Adv. Funct. Mater.}, 16:1086, 2006.

\bibitem{bianchi:ipcfirst}
E.~Bianchi, G.~Kahl, and C.~N. Likos.
\newblock {\em Soft~Matter}, 7:8313, 2011.

\bibitem{Blaak_2008}
C.~N. Likos, R.~Blaak, and A.~Wynveen.
\newblock {\em J. Phys.: Condens. Matter}, 20:494221, 2008.

\bibitem{Blaak_2012}
R.~Blaak and C.~N. Likos.
\newblock {\em J. Phys.: Condens. Matter}, 24:322101, 2012.

\bibitem{bianchi:2d2013}
E.~Bianchi, C.~N. Likos, and G.~Kahl.
\newblock {\em ACS~Nano}, 7:4647, 2013.

\bibitem{bianchi:2d2014}
E.~Bianchi, C.~N. Likos, and G.~Kahl.
\newblock {\em NanoLetters}, 14:3412, 2014.

\bibitem{noya:planes2014}
E.~G. Noya, I.~Kolovos, G.~Doppelbauer, G.~Kahl, and E.~Bianchi.
\newblock {\em Soft~Matter}, 10:8464, 2014.

\bibitem{noya:planes2015}
E.~G. Noya and E.~Bianchi.
\newblock {\em J.~Phys.:~Condens.~Matter}, in press, 2015.

\bibitem{yura:theory}
Y.~V. Kalyuzhnyi, E.~Bianchi, S.~Ferrari, and G.~Kahl.
\newblock {\em J. Chem. Phys.}, 142:14108, 2015.

\bibitem{silvano:30n}
S.~Ferrari, E.~Bianchi, Y.~V. Kalyuzhnyi, and G.~Kahl.
\newblock {\em J. Phys.: Condens. Matt.}, in press, 2015.

\bibitem{Kretzschmar_2013}
Z.~He and I.~Kretzschmar.
\newblock {\em Langmuir}, 29:15755, 2013.

\bibitem{Kretzschmar_2012}
Z.~He and I.~Kretzschmar.
\newblock {\em Langmuir}, 28:9915, 2012.

\bibitem{Kagome_2011}
Q.~Chen, S.~C. Bae, and S.~Granick.
\newblock {\em Nature}, 469:381, 2011.

\bibitem{Kraft_2009}
D.~J. Kraft, J.~Groenewold, and W.~K. Kegel.
\newblock {\em Soft Matter}, 5:3823, 2009.

\bibitem{vanOostrum_2015}
P.~D.~J. {van Oostrum}, M.~Hejazifar, C.~Niedermayer, and E.~Reimhult.
\newblock {\em J. Phys.: Condens. Matter}, in press, 2015.

\bibitem{Ravaine_2014}
A.~D{\'e}sert, C.~Hubert, A.~Thill, O.~Spalla, J.-C. Taveau, O.~Lambert,
  M.~Lansalot, E.~Bourgeat-Lami, E.~Duguet, and S.~Ravaine.
\newblock {\em Molec. Cryst. Liq. Cryst.}, 604:27, 2014.

\bibitem{Ravaine_2014bis}
T.~Blade, L.~Malosse, E.~Duguet, M.~Lansalot, E.~Bourgeat-Lami, and S.~Ravaine.
\newblock {\em Polym. Chem.}, 5:5609, 2014.

\bibitem{dlvo}
E.~J.~W. Verwey and J.~Th.~G. Overbeek.
\newblock {\em Theory of the stability of lyophobic colloids}.
\newblock Elsevier, Amsterdam, 1948.

\end{thebibliography}

\section{Appendix: IPCs with 3 patches}

To demonstrate the versatility of our approach we consider a particle of charge $Z_{\rm c}$ and diameter $2\sigma$, which is decorated by three patches of charges $Z_{\rm p_1}$, $Z_{\rm p_2}$, and $Z_{\rm p_3}$. The patches are located on the equatorial plane of the colloid, their respective centers-of-charges being separated by the distances $a_1$, $a_2$, and $a_3$ from the center of the colloid; connecting lines between a patch center and the particle center enclose throughout $120^{\degree}$.

The source term of Equation~(\ref{eq:poisson}) is given by
\begin{widetext}
\begin{eqnarray}
\rho(r,\theta,\varphi)=-Z_{\rm c}\delta({\bf{r}})&-&Z_{\rm p_1}\frac{1}{a_1^2}\delta(r-a_1)\delta(\theta-\frac{\pi}{2})\delta(\varphi) \\
                                            &-&Z_{\rm p_2}\frac{1}{a_2^2}\delta(r-a_2)\delta(\theta-\frac{\pi}{2})\delta\left(\varphi-\frac{2\pi}{3}\right) \nonumber \\
                                            &-&Z_{\rm p_3}\frac{1}{a_3^2}\delta(r-a_3)\delta(\theta-\frac{\pi}{2})\delta\left(\varphi-\frac{4\pi}{3}\right) \nonumber
\end{eqnarray}
\end{widetext}
leading to the following particular solution of the differential equation inside the colloid
\begin{widetext}
\begin{align}\label{eq:phi_in_3}
\Phi^{(1)}_{\rm part}(r,\theta,\varphi)&=
\frac{4\pi}{\varepsilon}\sum\limits_{{\ell=0}}^{\infty}\sum\limits_{m=-\ell}^{+\ell}\frac{1}{2\ell+1}\left[
    \frac{Z_{\rm p_1} a_1^{\ell}}{r^{\ell+1}}Y_{lm}^{*}\left(\frac{\pi}{2},0\right)+\frac{Z_{\rm p_2} a_2^{\ell}}{r^{\ell+1}}Y_{lm}^{*}\left(\frac{\pi}{2},\frac{2\pi}{3}\right) \right.  \\
&+ \left. \frac{Z_{\rm p_3} a_3^{\ell}}{r^{\ell+1}}Y_{lm}^{*}\left(\frac{\pi}{2},\frac{4\pi}{3}\right)\right]Y_{lm}(\theta,\varphi)+\frac{4\pi}{\varepsilon}Z_{\rm c}\frac{1}{r}Y_{00}^{*}Y_{00}. \nonumber
\end{align}
\end{widetext}
Combining expression~(\ref{eq:phi_in_3}) with the homogeneous solution given by expression~(\ref{eq:phi_in_hom}), we obtain the potential inside the colloid $\Phi^{(1)}(r,\theta, \phi)$. 

Outside the colloids, the solution of the Helmholtz equation~(\ref{eq:helmoltz}), $\Phi^{(2)} (r, \theta, \phi)$, is given by expression~(\ref{eq:phi_out_bis}).

Imposing on $\Phi_{\rm part}^{(1)}(r, \theta, \phi)$ and $\Phi_{\rm  part}^{(2)}(r, \theta, \phi)$ the boundary conditions~(\ref{eq:boundary}) at $r =\sigma$ leads to a set of linear equations for the unknown expansion coefficients that is solved numerically. In this way, we obtain the potential created by one three-patch IPC, $\Phi(r, \theta, \phi)$. It is worth noticing that for this patch decoration the axial rational symmetry is lost. 

\begin{figure}[h]
\includegraphics[height=5.5cm]{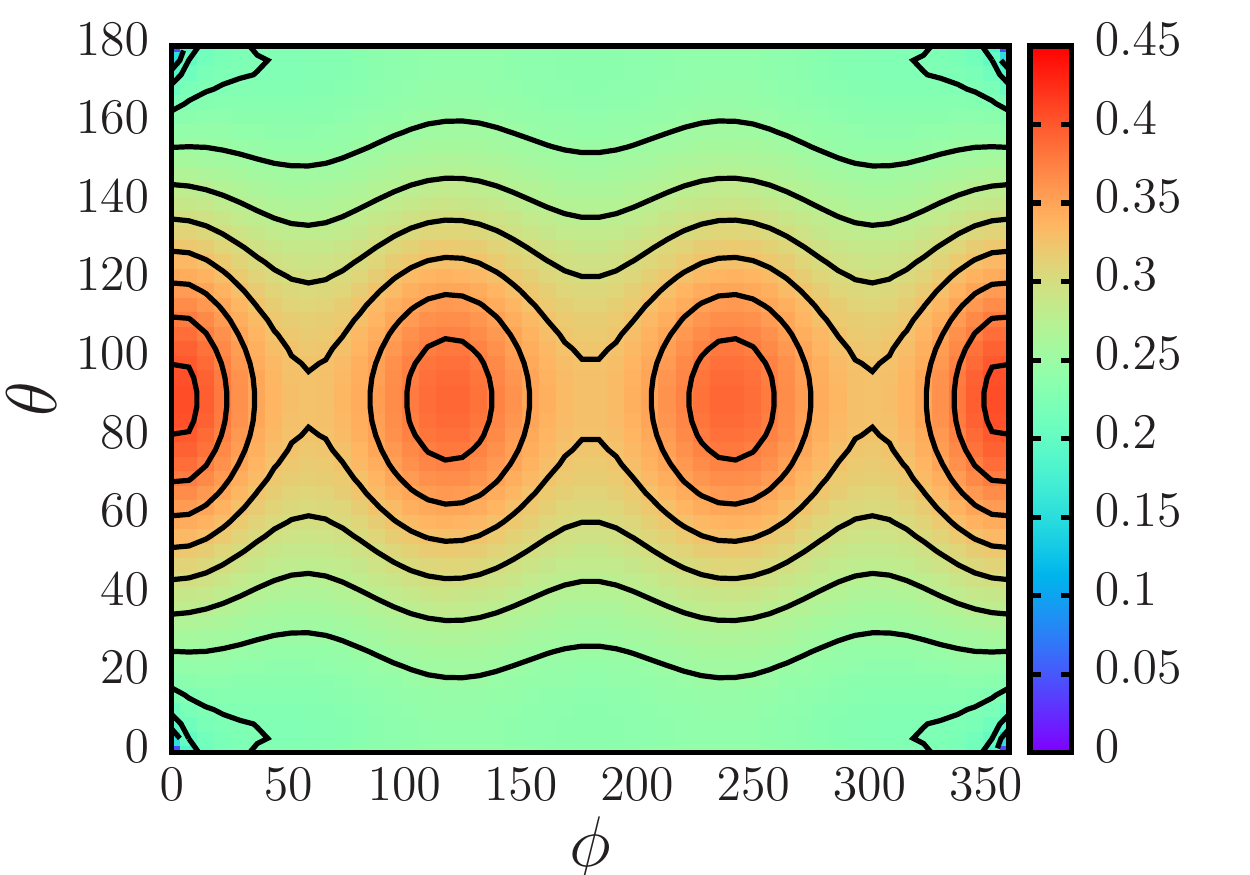}
\includegraphics[height=5.5cm]{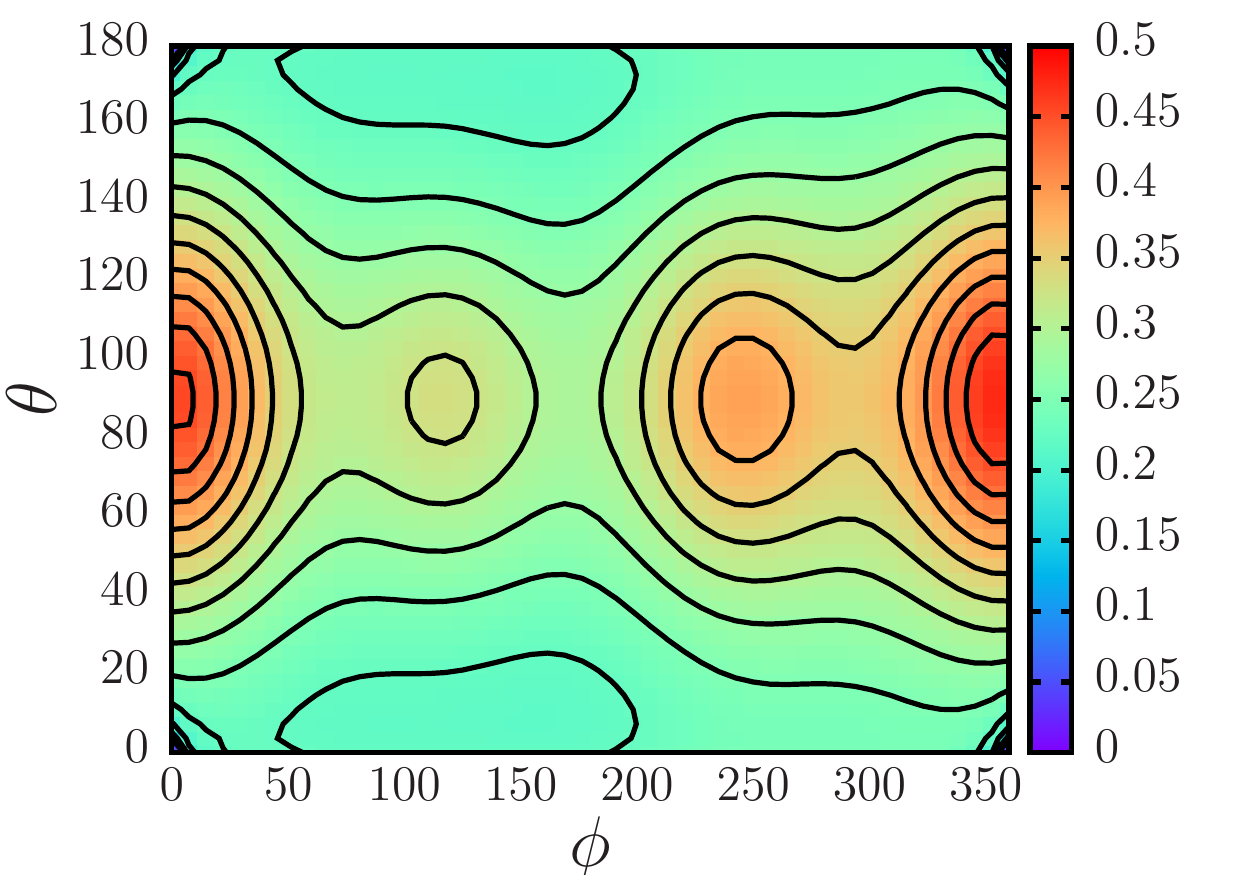}
\caption{The single particle potential, $\Phi(r, \theta, \phi)$, for an IPC with three identical (left panel) or differently charged (right panel) patches is reported at distance $r=\sigma$ from the particle center as a function of $\theta$ and $\phi$. For identical patches, the chosen parameters are: $Z_{\rm c}=-180$, $Z_{\rm p_1}=Z_{\rm p_2}=Z_{\rm p_3}=90$, $a_1=a_2=a_3=0.44\sigma$, $\kappa\sigma=5$, and $\epsilon=80$ (dielectric permittivity of water at room temperature). In the asymmetric case, the charges of the patches are chosen to be $Z_{\rm p_1}=110$, $Z_{\rm p_2}=70$, and $Z_{\rm p_3}=90$, while the other parameters are unchanged.}
\label{fig:singlepot_3}
\end{figure}

In Figure~\ref{fig:singlepot_3} we display $\Phi(r, \theta, \phi)$ for a IPCs with three identical patches (left panel) or three differently charged patches (right panel) at $r=\sigma$.  

The pair potential of two interacting IPCs is obtained by following the same procedure applied in the two patch case; the resulting interaction potential depends in a complex way on the center-to-center distance of the colloids and on their relative respective orientations.

\begin{figure}[h]
\includegraphics[height=9cm]{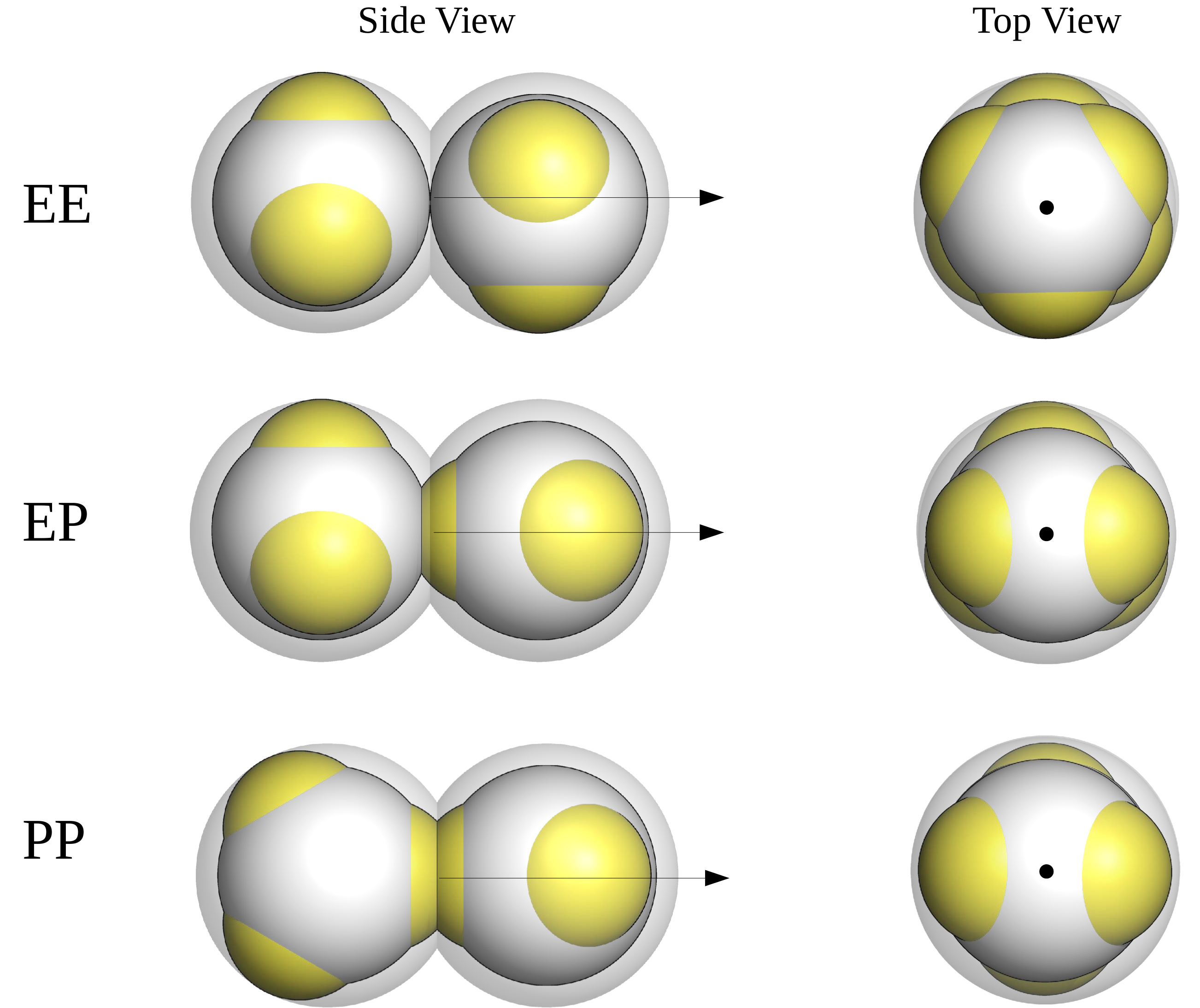}
\caption{Reference configurations for the mapping between the analytic and the coarse-grained pair potential of IPCs with three identical patches: characteristic configurations for the pure EE, EP and PP interactions are reported from top to bottom, as labeled; side views on the left, top views on the right. The second IPC can rotate with respect to the first one -- preserving pure EE, EP and PP interactions -- around an axis depicted as a black arrow in the side view or as a black dot in the top view.}
\label{fig:3patches}
\end{figure}

In parallel to the derivation of the Debye-H{\"uckel} potential, we design the coarse-grained model along the same lines traced in the body of the paper for the two patch case: we make an ansatz for $U$, similar to the one in Equation~(\ref{eq:U}), where the sum is now extended over all the possible pairings of the four interaction spheres. The evaluation of the overlap volumes $w_{ij}$ is straightforward given the relative position and orientation of the two IPCs, while the yet undetermined energy parameters $u_{ij}$ are again fixed by considering well-defined reference configurations. When the three patches are identical (in size and charge), then three $u_{ij}$-values must be determined, corresponding to the pure bare/bare, bare/patch, and patch/patch interactions. The most characteristic configurations needed to evaluate these pure interactions are depicted in Figure~\ref{fig:3patches}  (labeled again EE, EP, and PP, respectively). As the choice of the reference configurations is to some extent arbitrary, it is equally valid to consider rotations of one of the two IPCs that preserve the pure overlap between the selected interaction spheres. The rotation axis for the chosen configurations is depicted in Figure~\ref{fig:3patches}. The average value of the analytic contact energies of all equivalent configurations is then used as the reference for the mapping to the coarse-grained contact energies. 

It must be noted that the mapping procedure can be performed also when the three patches are different. In this case, the $u_{ij}$-values to be evaluated adds up to ten and thus ten reference configurations with the pure EE, EP and PP interactions must be considered: one EE configuration, three EP configurations and six PP configurations. Due to the reduced symmetry as compared to the identical patch case, averaging over all possible rotations of one IPC with respect to the other is advisable. The same rotation axis shown in Figure~\ref{fig:3patches} can be used.

\end{document}